\documentclass[aps,amsfonts,nofootinbib]{revtex4}
\begin{document}
\title{An Adiabatic Approximation to the Path Integral
for Relativistic Fermionic Fields}
\author{J. L. Cort\'es}
\email{cortes@unizar.es}
\affiliation{Departamento de F\'{\i}sica Te\'orica, Universidad de Zaragoza
50009, Spain}
\author{J. Gamboa}
\email{jgamboa@usach.cl}
\affiliation{Departamento de F\'{\i}sica,  Universidad de Santiago de Chile,
Casilla 307, Santiago 2, Chile, }
\author{J. Lopez-Sarrion}
\email{justo@dftuz.unizar.es}
\affiliation{ Departamento de F\'{\i}sica Te\'orica, 
Universidad de Zaragoza 
50009, Spain, and Departamento de F\'{\i}sica,  Universidad de Santiago
de Chile, Casilla 307, Santiago 2, Chile}
\author{S. Lepe}
\affiliation{Instituto F\'{\i}sica, Universidad Cat\'{o}lica de
Valpara\'{\i}so, Casilla 4059, Valpara\'{\i}so, Chile}
\begin{abstract}
A new approach to the path integral over fermionic fields, based on
the extension of a reformulation of the adiabatic approximation to
some quantum mechanical systems, is presented. A novel  non-analytic 
contribution to the efective fermionic action for a fermion field
coupled to a non-Abelian vector field is identified. The possible 
interpretation of this contribution as a violation of the decoupling theorem 
in Quantum Field Theory (QFT) is discussed. The generalization of the
approach to the case of finite temperature and density suggests the
possibility to apply it to the understanding of non-perturbative
properties in QFT and their dependence on temperature and density.
\end{abstract}
\maketitle
 \section{Introduction}

 The adiabatic approximation is one of the most important methods
 going beyond perturbation theory in quantum mechanics. In QFT,  the
 necessity of non-perturbative methods is clear in many cases (low
 energy limit of asymptotically free theories, high energy limit of
 infrared safe theories). Unfortunately, the attempts to translate the
 adiabatic approximation to QFT have been very limited and the main
 results are the identification of Wess-Zumino terms and anomalies as
 geometric phases \cite{stone}. The complexity of QFT (a quantum
 mechanical system with infinite degrees of freedom) has been an
 obstacle to the possible use of the adiabatic approximation as the
 starting point to an alternative to the perturbative expansion. In this
 paper we attempt to give a first step in this direction by
 considering the path integral over a relativistic fermionic field system. 
 In the next section we take as a starting point a very simple quantum
 mechanical system, a spin coupled to a time dependent magnetic field,
 and  the adiabatic approximation is reformulated in an appropriate
 way to be generalized to other systems. In section III, we point out
 the difficulties found when one tries a generalization of this
 reformulation in QFT. In section IV, a possible way 
 to circumvent these problems is presented and our proposal is applied
 to the path integral over a fermion field in the fundamental
 representation of $SU(2)$ coupled to a vector field in the adjoint 
 representation. The leading term in the adiabatic
 approximation is determined and its possible relation with
 non-perturbative properties of the theory are discussed at a
 qualitative level. In section V, a generalization of the
 results at finite density and temperature is obtained and finally in
 section VI a discussion of possible physical realizations of the
 adiabatic approximation in QFT is presented. 

 \section{A quantum mechanical example}

 In order to formulate the adiabatic aproximation in QFT, let us first
discuss it at the level of quantum mechanics in the most simple non-trivial
example, namely, a half-integer  spin ($j$) coupled to an external
magnetic field (${\vec B}(t)$) varying periodically in time (${\vec
  B}(T)={\vec B}(-T)$). Let us review the well known solution of this
problem \cite{berry}. One has a hamiltonian 
 \begin{equation} 
H \, = \, \sum_{\alpha,\beta} {\vec B}(t)\cdot {\vec J}_{\alpha\beta}
a^{\dagger}_{\alpha} a_{\beta}, 
\label{H1}
 \end{equation}
\noindent where $\alpha,\beta = -j, -j+1,...,j-1,j$ and $a$,
$a^{\dagger}$ are operators satisfying the (anti)commutation relations 
\begin{equation}
\lbrace a_{\alpha}, a^{\dagger}_{\beta}\rbrace \, = \delta_{\alpha\beta}.
\label{a,adagger}
\end{equation}

\noindent We want to determine the probability amplitude that the
system remains in the ground state ($|0>$) during time evolution, {\it
  i.e.}, the matrix element $<0(T)|0(-T)>$. In order to do this
calculation it is convenient to use at each time,  the direction of
the magnetic field as the  spin quantization axis to rewrite the
hamiltonian as 
\begin{equation}
H(t) \, = \, \sum_{m=-j}^{j} m B(t) a^{\dagger}_m(t) a_m(t).
\label{H2}
\end{equation}
The ground state of the system, $|0(t)>$ is the state satisfying the conditions
\begin{equation}
a_m(t) |0(t)> = 0 \:\:\: \mbox{for} \:\:\: m>0
{\hskip 2cm}
a^{\dagger}_m(t) |0(t)> = 0 \:\:\: \mbox{for} \:\:\: m<0.
\label{0(t)}
\end{equation}
\noindent In the adiabatic approximation one has two contributions to
the matrix element $<0(T)|0(-T)>$, one due to the energy of the ground
state 
\begin{equation}
E_0 (t) \, = \, \sum_{m<0} m B(t) \, = \, - \frac{(j+1/2)^2}{2} B(t),
\label{E0}
\end{equation}
\noindent and a contribution due to the time evolution of the phase of
the ground state $|0(t)>$ 
\begin{equation}
\int_{-T}^{T} dt <0(t)|i \partial_t|0(t)> = - {\cal I}m \int
d{\vec S}\sum_{k\neq 0}\frac{<0(t)|\nabla_{{\vec B}} H
  |k(t)> \wedge 
<k(t)|\nabla_{{\vec B}} H |0(t)>}{(E_k - E_0)^2}.
\end{equation}
\noindent Stoke's theorem has been applied to rewrite the time
integral as a surface integral in magnetic field space and the matrix
elements of the gradient of the hamiltonian can be directly read from
the matrix representation of the angular momentum operator. 

After some straightforward algebra one finds
\begin{equation}
\int_{-T}^{T} dt <0(t)|i \partial_t|0(t)> \, = \, \frac{(j+1/2)^2}{2} \Omega[{\hat B}],
\label{Berry}
\end{equation}
\noindent where
\begin{equation}
\Omega [{\hat B}] \, = \, \int d{\vec S}\cdot \frac{{\vec B}}{B^3},
\end{equation}
\noindent is the solid angle that ${\vec B}$ subtends. This is the
standard derivation of the adiabatic approximation including 
Berry's phase (\ref{Berry}) \cite{berry}.

\noindent Let us now see how these results can be rederived in an
alternative formulation based on a Grassmann path integral
representation of the evolution,  which will be useful in the
discussion of QFT in section IV. One can represent the operators $a$,
$a^\dagger$ by Grassmann variables $\psi$, ${\bar \psi}$ and the
probability amplitude that the system remains in the ground state by a
Euclidean Grassmann path integral, 
\begin{equation}
\lim_{T\to \infty} <0(T)|0(-T)> \, = \, \frac{\int d\psi d{\bar \psi}
  e^{-S}}{\int d\psi d{\bar \psi} e^{-S}|_{{\vec B}=0}} \, = \,
e^{-\Gamma[{\vec B}]}, 
\label{GammaB}
\end{equation}
\noindent where the Euclidean action for the fermionic system is
\begin{equation}
S \, = \, \int d\tau {\bar \psi} (\partial_{\tau} + {\vec J}.{\vec B}) \psi.
\label{Se}
\end{equation}
In order to simplify the action, one introduces new Grassmann
variables $c_m$, ${\bar c}_m$ through the expansion 
\begin{equation}
\psi = \sum_{m=-j}^{j} c_m f_m
{\hskip 2cm}
{\bar \psi}= \sum_{m=-j}^{j} f_m^{\dagger} {\bar c_m},
\end{equation}
\noindent where $f_m(t)$ are the eigenspinors defined by,
\begin{equation}
{\vec J}\cdot {\vec B} f_m \, = \, m B f_m.
\label{fmB}
\end{equation}
\noindent The action as a function of the new variables is given by,
\begin{equation}
S \, = \, \int d\tau \lbrace\sum _m \left[{\bar c}_m (\partial_{\tau}
  + m B) c_m \right] + \sum_{m,m^{'}}  {\bar c}_m i {\cal A}_{m,m^{'}}
c_{m^{'}} \rbrace, 
\end{equation}
\noindent where $i {\cal A}_{m,m^{'}} =  f_m^{\dagger} \partial_{\tau}
f_{m^{'}}$. In the adiabatic approximation, one neglects off-diagonal
terms 
($m\neq m^{'}$) and the path integral becomes a product of independent integrals
\begin{equation}
e^{- \Gamma^{(ad)}[{\vec B}]}\,  = \, \prod_{m=-j}^{j} \int dc_m
d{\bar c}_m e^{- \int_{-\beta/2}^{\beta/2} d\tau 
{\bar c}_m (\partial_{\tau} + i {\cal A}_m + m B) c_m},
\label{GammaadB}
\end{equation}
\noindent with ${\cal A}_m = {\cal A}_{m,m}$ and the limits on the
Euclidean time ($\tau =  i t$) incorporate finite temperature
($1/\beta$) effects in the imaginary time formalism \cite{kapusta}. 

\noindent The integral on each pair of variables ($c_m$, ${\bar
  c}_m$),  is a standard quantum mechanical determinant \cite{Dunne}  
\begin{equation}
\det [\partial_{\tau} + i {\cal A}_m(\tau) + m B(\tau)] \, = \, {\cal
  N} \, \cosh \left[\frac{\beta}{2} (m {\tilde B} + i {\tilde{\cal
      A}}_m\right], 
\label{detqm}
\end{equation}
\noindent where ${\cal N}$ is an infinite constant that will cancel in
the ratio of Grassmann integrals in (\ref{GammaB}) and  we have
introduced the notation 
\begin{equation}
{\tilde f} \, = \, \frac{1}{\beta} \int _{-\beta/2}^{\beta/2} d\tau \, f(\tau).
\label{tilde}
\end{equation}
\noindent In the zero temperature limit the quantum mechanical
determinants (\ref{detqm}) become exponentials and the effective
fermionic action in the adiabatic approximation takes the simple form 
\begin{equation}
- \Gamma^{(ad)}[{\vec B}] \, = \, - \frac{i}{2} \sum_m \left( |m| \int
  dt B(t) + \frac{m}{|m|} \int dt {\cal A}_m(t) \right), 
\label{GammaadB2}
\end{equation}
\noindent where the first term can be recognized as the dynamical
phase $-i\int dt E_0(t)$ with $E_0$ the energy of the ground state of
the quantum mechanical system (\ref{E0}). The second term in
(\ref{GammaadB2}) reproduces Berry's phase as one can show by using
once more Stokes theorem and the definition of $f_m$ in (\ref{fmB}),
{\it i.e.} 
\begin{equation}
\int dt i {\cal A}_m (t) \, = \, \int dt f_m^{\dagger} \partial_t f_m
\, = \, - i m \Omega[{\hat B}], 
\label{Am}
\end{equation}
\noindent with $\Omega[{\hat B}]$ the solid angle that the magnetic
field subtends in its evolution. 

The adiabatic approximation to this simple quantum mechanical system
in the Grassmann path integral representation,  will reappear as an
ingredient in some approximation to a QFT with fermionic fields as we
will see later. Also this reformulation of the adiabatic approximation
is interesting because it allows to go beyond the zero temperature
limit by using the quantum mechanical determinants in (\ref{detqm}). 

 \section{Formal direct approach and its difficulties}

 The purpose of this section is to introduce a direct extension to QFT
 of the reformulation of the previous section. Before doing that we
 will show the problems of a direct implementation of the adiabatic
 approximation. 

 The most natural way to formulate the adiabatic approximation and the
 related Born-Oppenheimer approach in QFT is based on the use of the
 Schr\"odinger representation,  where the wave function of quantum
 mechanics is replaced by a functional in the space of field
 configurations. After this replacement,  all the standard results of
 the adiabatic quantum mechanical expansion can be applied directly in
 QFT \cite{Ralston}. This formulation gives a new perspective of the
 anomaly in chiral gauge theories which appears as a geometric phase
 in the space of gauge field configurations related to the gauge
 non-invariance of the phase of the fermionic Fock states
 \cite{Niemi}. Unfortunately,  a quantum mechanical system with 
infinite degrees of freedom is too complicated to go beyond the study
 of a few topological properties and the adiabatic approximation
 remains as a reformulation of the theory at a formal level. 

An alternative way to implement the adiabatic formulation is based on
a direct use of the reformulation of the quantum mechanical example of
the previous section in a QFT system. For definiteness,  let us
consider the action in Euclidean space 
for a fermionic field $\Psi$ coupled to a vector field $A_\mu$,

\begin{equation}
{\cal S} = \int d^D x \, \left\lbrace {\bar \Psi} \gamma_\mu
\left(\partial_\mu + i e A_\mu\right) \Psi \right\rbrace.
\label{S1}
\end{equation}

 Using the decomposition
 \begin{equation}
 \Psi = \left(\matrix{\Psi_R\cr \Psi_L}\right)
  {\hskip 2cm}
 {\bar \Psi} = \left( \Psi^{\dagger}_L \;\; \Psi^{\dagger}_R \right),
 \label{psiLR}
 \end{equation}
 \noindent for the fermionic field in a representation where all gamma
 matrices are off-diagonal, one has 
 \begin{eqnarray}
 {\cal S} = \int d^4 x \, &\lbrace &\Psi^{\dagger}_L \left(\partial_4 +
 ieA_4\right) \Psi_L
 + \Psi^{\dagger}_R \left(\partial_4 + ieA_4\right) \Psi_R 
\nonumber \\
 &+& \Psi^{\dagger}_L {\vec \sigma}.\left(-i{\vec \nabla} + e{\vec A}\right)
 \Psi_L - \Psi^{\dagger}_R {\vec \sigma}.\left(-i{\vec \nabla} +
 e{\vec A}\right) \Psi_R \rbrace.
 \label{S2}
 \end{eqnarray}

 Following the steps of the previous section,  we introduce the
 eigenfunctions $\Phi_n ({\vec x})$ 
\begin{equation}
\left[{\vec \sigma}.\left(-i{\vec \nabla} + e{\vec A}\right)\right]
\;  \Phi_n ({\vec x}) = \epsilon_n \Phi_n ({\vec x}).
\label{phi_n}
\end{equation}
\noindent These eigenfunctions and the eigenvalues $\epsilon_n$ are,
in fact,  functionals of the vector field at a given time and a more
precise notation for them is $\Phi_n[{\vec A}(\tau)]({\vec x})$ and
$\epsilon_n[{\vec A}(\tau)]$.

Next step is to use the decomposition of the fermionic fields in terms
of the eigenfunctions in (\ref{phi_n}),
\begin{equation}
\Psi_L({\vec x},\tau) =  \sum_n c_{L_n} (\tau) \;
 \Phi_n[{\vec A}(\tau)]({\vec x}) {\hskip 1cm}
\Psi_R({\vec x},\tau) =  \sum_n  c_{R_n} (\tau) \;
 \Phi_n[{\vec A}(\tau)]({\vec x})
\label{c}
\end{equation}
\begin{equation}
\Psi_L^{\dagger}({\vec x},\tau) = \sum_n c_{L_n}^{\dagger} (\tau)   \;
 \Phi_n^{\dagger}[{\vec A}(\tau)]({\vec x}) {\hskip 1cm}
\Psi_R^{\dagger}({\vec x},\tau) =  \sum_n  c_{R_n}^{\dagger} (\tau) \;
 \Phi_n^{\dagger}[{\vec A}(\tau)]({\vec x}).
\label{cdagger}
\end{equation}

Using the orthogonality of the eigenfunctions $\Phi_n$, the action
takes a compact form in terms of the (Grassman) coeficients $c_L$, $c_R$
\begin{eqnarray}
 {\cal S} \, = \int d\tau &\lbrace & \sum_n \left[c_{L_n}^{\dagger} (\tau)
\partial_{\tau} c_{L_n} (\tau) +
 c_{L_n}^{\dagger} (\tau)\epsilon_n c_{L_n} (\tau) +
i  c_{L_n}^{\dagger} (\tau) {\cal A}_n c_{L_n} (\tau)\right]
\nonumber \\
&+& \sum_n  \left[c_{R_n}^{\dagger} (\tau)
\partial_{\tau} c_{R_n} (\tau) -
 c_{R_n}^{\dagger} (\tau)\epsilon_n c_{R_n} (\tau) +
i  c_{R_n}^{\dagger} (\tau) {\cal A}_n c_{R_n} (\tau)\right]
\nonumber \\
&+& \sum_{n\neq m} i \left[c_{L_n}^{\dagger} (\tau) {\cal A}_{nm} c_{L_m} (\tau) +
c_{R_n}^{\dagger} (\tau) {\cal A}_{nm} c_{R_m} (\tau)\right] \rbrace,
\label{S3}
\end{eqnarray}
\noindent where we have introduced the connection ${\cal A}_n$,
\begin{equation}
{\cal A}_n [A(\tau)] = \int d{\vec x} \;
\Phi_n^{\dagger}[{\vec A}(\tau)]({\vec x}) \left(-i \partial_{\tau} + eA_4\right)
\Phi_n[{\vec A}(\tau)]({\vec x})
\label{calA_n}
\end{equation}
\noindent and ${\cal A}_{nm}$ for $n\neq m$,
\begin{equation}
{\cal A}_{nm} [A(\tau)] = \int d{\vec x} \;
\Phi_n^{\dagger}[{\vec A}(\tau)]({\vec x}) \left(-i \partial_{\tau} + eA_4\right)
\Phi_m[{\vec A}(\tau)]({\vec x})
\label{calA_nm}
\end{equation}

In the adiabatic approximation one neglects the off-diagonal terms
($n\neq m$) and then one has infinite copies of quantum mechanical
systems each of them similar to the one discussed in section
II. However for a general vector field configuration the spectrum
($\epsilon_n$) will be continous and the difference of energy levels
can be arbitrarily small rendering the adiabatic expansion  out of
control. Besides that, the eigenvalues $\epsilon_n$ and
eigenfunctionals $\Phi_n$ are not known, except for very special
choices of the vector field, and the formulation remains
once more at a formal level. 

 \section{ A new approach and its application to an $SU(2)$ gauge theory}

 The only way we have found to use the reformulation of the adiabatic
 approximation to get a useful expansion in QFT,  is based on the
 introduction of variables independently at each point. In order to do
 that,  one has to select an operator at each point and use its
 eigenfunctions in the  expansion of some of the fields at this
 point. We can then identify two ingredients in the formulation of the
 new approach. The first one is a separation of the fields in two
 sets, one of them corresponding to the spin degrees of freedom of the
 quantum mechanical example. The second one  is the choice of the
 operator at each point whose eigenfunctions are used to introduce new
 variables for the spin-like fields. 

 Several requirements constraint the ambiguities in this two
 ingredients. The action should be quadratic in the spin-like fields
 either directly or after the introduction of appropriate auxiliary
 fields. The expression for some of the terms in the action as a
 function of the new variables should be as simple as possible. The
 contribution from the remaining terms in the action (including the
 space derivatives) as well as the corrections to the \lq\lq
 adiabatic" approximation (off-diagonal contributions in the new
 variables) should be small. The search of a good set of fields and
 local operators defining the new variables has to be done, however
 case by case. The usefulness of the approach will be established if
 one finds examples where all these requirements are satisfied
 yielding to a dominant contribution with interesting results. 

In order to illustrate our approach let us consider the Lagrangean of
a fermionic system coupled to a vector field 
 \begin{equation}
 {\cal L} = {\bar \Psi} i \gamma^{\mu} \partial_{\mu} \Psi -
 g {\bar \Psi} \gamma^{\mu} A_{\mu}^aT^{a} \Psi - m {\bar \Psi} \Psi,
 \label{Ls1}
\end{equation}
 \noindent where $T^a$ are the generators of $SU(2)$ acting on the
 fermionic fields in the fundamental representation. It is convenient
 to use the Dirac representation for the $\gamma$ matrices with 
 \begin{equation}
 \gamma^0 = \left(\matrix{ I & 0\cr 0 & - I} \right)
 \label{gamma0}
\end{equation}
 \noindent and the decomposition in bispinors of the Dirac field
 \begin{equation}
 \Psi = \left(\matrix{\varphi \cr \chi}\right).
 \label{Psi}
\end{equation}
 We neglect for a moment, the terms proportional to the space
 components of the vector field (${\vec A}^a$) and to space
 derivatives of the fermionic field. The remaining terms take a simple
 form if we use the eigenvectors $f_{\pm}$ of the operator ${\hat
 A}_0^a T^a$ 
  \begin{equation}
 \left({\hat A}_0^a T^a\right) \, f_{\pm} \, = \pm \frac{1}{2} \, f_{\pm},
 \label{f}
\end{equation}
\noindent where we have used the parametrization $A_0^a = A_0 {\hat
  A}_0^a$ with $\sum_a {\hat A}_0^a {\hat A}_0^a =1$. 
With these eigenvectors,  one can introduce the new fermionic variables
$\varphi_{n,i}$, $\chi_{n,i}$
\begin{equation}
\varphi \, = \sum_{n=\pm} \sum_{i=1,2} \varphi_{n,i} f_{n,i},
\label{varphi}
\end{equation}
\begin{equation}
\chi \, = \sum_n \sum_{i=1,2} \chi_{n,i} f_{n,i},
\label{chi}
\end{equation}
where the bispinors $f_{n,i}$ are given by
\begin{equation}
f_{n,1} \, = \, \left(\matrix{ f_n \cr 0}\right),
{\hskip 2cm}
f_{n,2} \, = \, \left(\matrix{ 0 \cr f_n }\right).
\label{fni}
\end{equation}
\noindent Note that the new fermionic variables have been introduced
independently at each point in space. We then have Grassmann variables
$\varphi_{n,i}$, $\chi_{n,i}$ at each space-time point. 

In the new representation for the fermionic variables,  one has
\begin{equation}
\Psi^{\dagger} \partial_{\tau} \Psi \, =
\sum_{n,i} \left[\varphi_{n,i}^{\dagger} \partial_{\tau} \varphi_{n,i} \, + \,
\chi_{n,i}^{\dagger} \partial_{\tau}\chi_{n,i} \right] \, + \,
\sum_{n,n^{'},i} \left[\varphi_{n,i}^{\dagger} \varphi_{n^{'},i} \, + \,
\chi_{n,i}^{\dagger} \chi_{n^{'},i} \right] \,  i {\cal A}_{n,n^{'}}
\label{partialtau}
\end{equation}
\noindent with $i {\cal A}_{n,n^{'}} = f_n^{\dagger} \partial_{\tau}
f_{n^{'}}$ and $\tau= i t$ is the Euclidean time. 

For the interaction term one has
\begin{equation}
- g \Psi^{\dagger} A_0^{a} T^{a} \Psi \, = \, - \sum_n g_n A_0 \sum_i
\left[\varphi_{n,i}^{\dagger} \varphi_{n,i} \, + \,
\chi_{n,i}^{\dagger} \chi_{n,i} \right],
\label{int0}
\end{equation}
\noindent with $g_{\pm} = \pm \frac{g}{2}$.

The Lagrangian density before including space derivatives and the
space components of the vector field as a function of the new
variables is 
\begin{eqnarray}
{\cal L} (A_0^a) \,& = & \, \sum_{n,i} \left[\varphi_{n,i}^{\dagger}
\left(\partial_{\tau}- g_n A_0 - m + i {\cal A}_n\right) \varphi_{n,i}
\, + \, \chi_{n,i}^{\dagger}
\left(\partial_{\tau}- g_n A_0 + m + i {\cal A}_n\right) \chi_{n,i} \right]
\nonumber \\ \,& + & \,
\sum_{n\neq n^{'},i} \left[\varphi_{n,i}^{\dagger} \varphi_{n^{'},i} \, + \,
\chi_{n,i}^{\dagger} \chi_{n^{'},i} \right] \, i {\cal A}_{n,n^{'}}.
\label{L0}
\end{eqnarray}

\noindent We then have at each point,  a generalization of the quantum
mechanical example in section II with $A_0^a$ playing the role of the
magnetic field ${\vec B}$ (the direction in the internal $SU(2)$ space
is the analog to the orientation of the magnetic field) and four
instead of one spin variable 
(corresponding to the four components of the Dirac spinor). Then,  the
energies associated to the new variables are 
\begin{equation}
E(\varphi_{n,i}) = g_n A_0 + m,
{\hskip 3cm}
E(\chi_{n,i}) = g_n A_0 - m.
\label{E}
\end{equation}

The corrections to the adiabatic approximation due to the off-diagonal
terms ($n\neq n^{'}$) in (\ref{L0}) involve levels separated by a gap
$g A_0$. Then, the adiabatic approximation will be justified when the
product of the coupling and  the time component of the vector field is
large compared to the time derivative of its orientation (${\dot {\hat
    A}}_0^a$).  

With respect to the consistency of treating the space derivatives and
space components of the vector field as a correction, we have a sum of
two contributions, 
\begin{eqnarray}
\Psi^{\dagger} {\vec \alpha}.(-i\vec{\nabla}&+&g\vec{A}^aT^a) \Psi \,=\,
\sum_{n,i,i^{'}} \left[\varphi_{n,i}^{\dagger} (-i\vec{\nabla}) \chi_{n,i^{'}}
\, + \, \chi_{n,i}^{\dagger} (-i\vec{\nabla}) \varphi_{n,i^{'}}\right] \,
\left(f_{n,i}^{\dagger} {\vec \sigma} f_{n,i^{'}}\right) \nonumber \\ &+& \,
\sum_{n,n^{'},i,i^{'}} \left[\varphi_{n,i}^{\dagger} \chi_{n^{'},i^{'}} \, +
\chi_{n,i}^{\dagger} \varphi_{n^{'},i^{'}}\right] \,
\left(f_{n,i}^{\dagger} {\vec \sigma}.(-i {\vec \nabla}+g\vec{A}^aT^a) f_{n^{'},i^{'}}
\right).
\label{nabla}
\end{eqnarray}
\noindent The first term is a non-diagonal contribution between energy
levels separated by $2m$. Since the eigenvectors $f_n$ depend only on
the direction in the internal space of $A_0$, then  the corrections
due to these terms will be proportional to $\vec{\nabla}{\hat
  A}_0^a/2m$. The second term has non-diagonal contributions between
energy levels separated by $2m$, $2m+gA_0$ or $|2m-gA_0|$ and there
are two types of terms, ones proportional to $\vec{\nabla}{\hat
  A}_0^a$ and the others proportional to $g\vec{A}^a$. From these
simple arguments one can see what are the conditions on the vector
field and the fermion mass  in order to treat (\ref{nabla}) as a small
correction to (\ref{L0}). It should be  
noted that considering spacial derivatives as corrections does not mean that
we are making use of the usual derivative expansion method~\cite{WZ}.

Then, in the approximation where we keep only the terms diagonal in
(\ref{L0}) and using the result (\ref{GammaadB2}-\ref{Am}) we have a
new approximation to the effective fermionic action, 

\begin{eqnarray}
\Gamma_{ad} \,& = & \, \int d^3{\vec x} \, \sum_n \,
sgn (g_n A_0 + m)\left[\int d\tau ( g_n A_0 + m ) +
\frac{g_n}{g} i \Omega[{\hat A}_0^a]\right] \nonumber \\ \, & + & \,
\int d^3{\vec x} \, \sum_n \,
sgn (g_n A_0 - m)\left[\int d\tau ( g_n A_0 - m ) +
 \frac{g_n}{g} i \Omega[{\hat A}_0^a]\right],
\label{Gamma_ad}
\end{eqnarray}
\noindent where $\Omega[{\hat A}_0^a]$ is the solid angle that ${\hat
  A}_0^a$ subtends in internal space in its time evolution. 

We can distinguish two different regions. In the weak coupling region
($g A_0 \ll 2 m$) there is a cancellation of contributions from the
two sets of new variables $\varphi_{n,i}$ and $\chi_{n,i}$ and
$\Gamma_{ad}$ reduces to a constant. On the other hand, in the strong
coupling region ( $g A_0 \gg 2 m$)  the effective fermionic action in
the adiabatic approximation is 
\begin{equation}
\Gamma_{ad}^{(s)} \, = \, 2 g  \int d\tau d^3 {\vec x} \, A_0  \, + 2
i \Omega[{\hat A}_0^a]. 
\label{Gammas}
\end{equation}
\noindent The presence of a contribution in the effective fermionic
action which dissapears in the weak coupling region,  suggest a
possible relation between the presence of such contribution and the
non-pertubative properties of the theory. 

One should note that the limit $m\to \infty$ --for fixed $g A_0$-- corresponds
to the weak coupling region,  where the new non-perturbative contribution 
dissapears as expected from the decoupling theorem \cite{appel}. On the other 
hand for an arbitrarily large mass,  one can be in the strong coupling limit  
if one has a sufficiently large coupling $g$ and/or vector field $A_0$ and 
then the non-perturbative trace of the fermionic system remains.

\section{Finite density and temperature effects}

The generalization of the adiabatic approximation to the $SU(2)$ gauge
theory in the case of finite density is trivial. All one has to do is
to include the chemical potencial ($\mu$) through a term $\mu {\bar
  \Psi} \gamma^0 \Psi$ in the Lagrangian which modifies the energies
associated to the new variables 
\begin{equation}
E(\varphi_{n,i}) = g_n A_0 + m + \mu,
{\hskip 3cm}
E(\chi_{n,i}) = g_n A_0 - m + \mu.
\label{Emu}
\end{equation}
\noindent The energy differences are not modified and  all the
estimates of the corrections due to space derivatives and the space
components of the vector field are not changed. 

The effective fermionic action in the adiabatic approximation is now,

\begin{eqnarray}
\Gamma_{ad} (\mu)&& =  \int d{\vec x} \, \sum_n \,
sgn (g_n A_0 + m + \mu)\left[\int dt ( g_n A_0 + m + \mu) +
\frac{g_n}{g}  i \Omega[{\hat A}_0^a]\right] \nonumber \\ +&&
\int d{\vec x} \, \sum_n \,
sgn (g_n A_0 - m + \mu)\left[\int dt ( g_n A_0 - m + \mu) +
 \frac{g_n}{g}  i \Omega[{\hat A}_0^a]\right].
\label{Gamma_admu}
\end{eqnarray}

We consider $\mu >0$ for definiteness. In this case,  we can consider
three different regions.  The first one corresponds to $gA_0 <
2|m-\mu|$ where once more there is a cancellation of contributions and
one has a trivial adiabatic effective action as in the weak coupling
region of the case without chemical potencial. There is also an analog
of the strong coupling region where $gA_0 > 2|m+\mu|$ and the
adiabatic effective action is 
(\ref{Gammas}). Finally,  there is an intermediate region, $2|m-\mu| <
gA_0 < 2|m+\mu|$ where there is a cancellation of the contributions of
half of the new fermionic variables and the result for the effective
action is 
\begin{equation}
\Gamma_{ad}^{(i)} \, = \,  g  \int d\tau d^3 {\vec x} \, A_0  \, +  i
\Omega[{\hat A}_0^a] + \mbox{constant}. 
\label{Gammai}
\end{equation}

Once more we can discuss the decoupling of the fermion degrees of
freedom. There are once more cases with arbitrarily large mass and
density where for sufficiently large $gA_0$ a fermionic signal
remains. On the other hand for fixed mass and $gA_0$ the fermion
decouples in the limit $\mu \to \infty$. This shows that the
non-perturbative properties of the non-abelian gauge theory related to
the presence of a non-trivial adiabatic effective action  dissapear in
the infinite density limit. 

It is also very easy to generalize the adiabatic approximation to the
case of finite temperature. As we have seen in section II,  the
modification of the fermionic integral for each quantum mechanical
system is very simple and the effective fermionic action in the
adiabatic approximation at finite temperature (and also including a
chemical potencial) is given by 
\begin{eqnarray}
- \Gamma_{ad} (\mu,\beta)\,& = & \, \int d{\vec x} \, \sum_n \,
\ln \cosh \left[\int_{-\beta/2}^{\beta/2} d\tau ( g_n A_0 + m + \mu) +
i \Omega[{\hat A}_0^a]\right] \nonumber \\ \, & + & \,
\int d{\vec x} \, \sum_n \,
 \ln \cosh \left[\int_{-\beta/2}^{\beta/2} d\tau ( g_n A_0 - m + \mu) +
i \Omega[{\hat A}_0^a]\right].
\label{Gamma_admuT}
\end{eqnarray}
\noindent where $\beta$ is the inverse of the temperature. If
one takes the high temperature limit, $\beta \to 0$ for fixed ($\mu$,
$m$, $gA_0$), then the adiabatic effective action is proportional to
$\beta^2$ and then the new non-perturbative  contribution dissapears. 

\section{Discussion}

The result for the effective fermionic action in the adiabatic
approximation (\ref{Gamma_ad}) is not gauge invariant neither Lorentz
invariant. This should not be a surprise. A similar situation happens
in the simplest quantum mechanical example of the spin coupled to a
magnetic field~\cite{Jac} where each term in the adiabatic expansion is not
separately rotational invariant, the variation of each term beeing
cancelled by that of subsequent terms in the expansion. This makes
difficult to understand how one can find a physical situation where
the adiabatic approximation to the effective fermionic action
(\ref{Gamma_ad}) can give a consistent first order approximation to a
relativistic QFT with fermionic fields. 

One possibility is to consider a gauge field theory, identifying the
vector field $A_{\mu}^a$ with the gauge field of the $SU(2)$ gauge
theory. In that case one should add a term in the action depending on
the dynamical gauge field. The Lorentz and gauge invariance of the
theory is inconsistent with the adiabatic approximation; in fact this
could have been anticipated because by applying an arbitrary gauge transformation
on a vector field whose space components are sufficiently small and
whose time component orientation in internal space varies sufficiently
slowly to justify the approximation one gets a gauge field
configuration where the approximation is not justified. The only way
to look for a consistent realization of the adiabatic approximation in
a gauge field theory is by considering the vector field as the gauge
field satisfying an appropriate noncovariant gauge fixing
condition. It is not possible with our present understanding of gauge
field theories at the nonperturbative level to check wether there is a
situation where a gauge fixing condition can be found such that the
relevant gauge field configurations in that gauge satisfy all the
conditions to justify the adiabatic approximation to the effective
fermionic action considered in this work. All one can do is to explore
the consequences of the assumption that this is the case.           
In all cases where the effective action in the adiabatic approximation
is non trivial one has a term proportional to $\int d\tau d^3 {\vec x}
A_0$ and then, in order to have a finite action, the gauge field $A_0$
should be concentrated in a finite region in space-time. This fact,
together with the disappearance of the new contribution in the high
temperature or high density limits suggests a relation of the presence
of a non-trivial adiabatic approximation and the confinement in the
non-abelian gauge theory.  Note that if this relation holds,  then the
confinement at low energies would be due to the presence of heavy
quarks which do not decouple as one would naively expect due to
non-perturbative effects.

Other possibilities for a physical realization of the adiabatic
approximation to the effective fermionic action could correspond to a
situation where the vector field is not a dynamical field but a
background field which parametrizes some of the nonperturbative
properties of the vacuum of QFT or an auxiliary field introduced to
linearize a fermion self-coupling. In those cases the Lorentz and
gauge non invariance of the result inherent to the adiabatic
approximation should be related with the details of the introduction
of the background or auxiliary field.  
 
\section{Acknowledgments} 
This work has been partially supported by the
grants 1050114, 7010516  Fondecyt- Chile, DI-PUCV
No. 123.771/2004 (SL),  by AECI (Programa de
Cooperaci\'on con Iberoam\'erica) and by MCYT (Spain), grant
FPA2003-02948. JLS thanks to MECESUP- 0108 and the Spanish Ministerio
de Educaci\'on y Cultura for support. 

\end{document}